\documentclass[12pt]{article}
\usepackage{amsfonts,amsmath,amssymb}
\usepackage{graphicx}
\usepackage{tikz}

\def\be{\begin{equation}}
\def\ee{\end{equation}}
\def\bea{\begin{eqnarray}}
\def\eea{\end{eqnarray}}

\def\half{{\textstyle {1\over 2}}}

\def\ul{\underline}

\begin{document}

\thispagestyle{plain}

\title{\bf\Large Feynman's proof of the commutativity of the Calogero integrals of motion}

\author{Alexios P. Polychronakos\\
{\tiny  .}\\
Department of Physics, City College of New York, NY 10038\\
{\it apolychronakos@ccny.cuny.edu}}

\noindent

\maketitle

\begin{abstract}
During his last year of life Feynman became interested in integrable models. In the course of his study of one dimensional particles with inverse square-type potentials (the Calogero class of models) he came up with a field theory-inspired proof of the commutativity of the integrals of motion of the system, which remained unpublished. I present an exact transcript of the proof together with a scan of the first page of Feynman's original manuscript and some historical background.
\end{abstract}
\vskip 0.3cm
{\centerline{{\it In memory of Richard P.\ Feynman}}}

\section{ Introduction }

2018 marks the 100$^{th}$ anniversary of the birth and 30 years since the passing of Richard P. Feynman,
an iconic figure of 20$^{ th}$ century science. Feynman dominated theoretical physics with his creativity, insights and style,
but also projected a sense of excitement about science and life in general, an infectious 
(and somewhat mischievous) {\it joie de vivre}. Altogether, Feynman single-handedly created the image of the playful, irreverent physicist relying heavily on his smarts and intuition, and achieved cult status well beyond the physics community.

I interacted with Feynman during what was to be his last year of life and my last year of graduate studies at Caltech. 
I was left with fond memories of exciting conversations and some of his personal notes and manuscripts. These include a proof of the commutativity of the Calogero integrals of motion,
written by Feynman as a letter to Bill Sutherland. The proof is interesting in that it uses a field theory-inspired technique
and involves some of Feynman's signature tricks, shorthands and elliptical thinking. Further, it is one of Feynman's very
last pieces of work.

On the occasion of Feynman's 100$^{th}$ anniversary, I would like to share this proof with the physics community.

\section{Historical background}

During the early to mid 1980s Feynman's interests were mainly in quantum computation, a topic he pioneered. In 1983 he taught a course on complexity and quantum limitations in computing, consisting of lectures by himself and seminars by guest speakers. In early 1986 the Challenger disaster occurred and Feynman became part of the investigation panel, leading to some interesting stories.

In the summer of 1986 Feynman suffered a recurrence of his cancer and had to undergo therapy, which kept him out of campus for much of the fall. When he recovered, he felt that he wanted to learn something new and turned to a subject spearheaded by his lifelong friend, Hans Bethe: integrable models and the Bethe Ansatz. He read, learned, absorbed and digested the material, and was ready to give it his own spin and intuition and enjoy it in his preferred way: by teaching it.
He gave a seminar, which spread into two sessions, where he exposed everything he knew about the subject in his inimitable way and described some of his ideas and hopes on where integrability might prove fruitful: QCD fragmentation, solving gauge theories by identifying ``good" variables like in integrable models.

After the lectures Feynman invited graduate students interested in the subject to meet with him regularly and discuss, and occasionally went with them to lunch at the student cafeteria, the infamous Chandler. Many flocked (I remember Michael Douglas, Olivier Espinosa, Arun Gupta, Sandip Trivedi; maybe others too) but eventually Arun, Sandip and myself remained.

One of the topics discussed was the inverse-square particle model (what we now call the Calogero model \cite{Calo}). Feynman had read a review by Sutherland and was fascinated by the model's many interesting properties (its ``surprises" he would say) and wanted to explore it further. In particular, he wanted to find an intuitive proof of the model's integrability. In this quest, he devised a field theory-based proof (and, of course, his own notation) which he described to us. He sent his proof to Sutherland with the comment ``I am disappointed in the proof... I learn nothing, no real clue as to why all this works, and what it means."

\section{Remarks about the proof}

There are at least two versions of the proof and the manuscript, and several versions of the printed letter, 
dating from mid- to end of May 1987. After writing the initial version, proving the commutativity of two
arbitrary integrals, Feynman switched to a generating function approach, much along the lines of introducing 
a spectral parameter. This also allowed him to dispense with an even more drastically simplified
notation that became unnecessary in the new proof. Feynman handwrote his proof and gave it to his secretary, the 
redoubtable and unforgettable Helen Tuck, who typed it, texed it and passed it to Feynman and myself for corrections.

After reviewing the first proof of the letter on May 26 Feynman performed several corrections
and instructed Helen to omit part of the old proof and insert his new proof based on the generating function. 
There are also some claims in the original manuscript that
were removed from the final version, presumably because they did not pan out or were not fully proven. The
new manuscript was corrected again at least twice, once by Feynman and once by myself. (My edits consisted mainly of fixing typos or misformatted equations and correcting the spelling of some words, especially non-English names like
Weierstrass.) The final version of the paper, addressed ``Dear Bill" rather than the formal ``Dear Mr Sutherland" of the original manuscript, was typed on May 29. An additional version, identical to the May 29 one,
was also produced on February 25, 1988, ten days after Feynman's passing, apparently printed by Helen to be included
in his files. All handwritten and printed versions are in my archive.

I will present below the final version of the proof (the May 29, 1987 version). In producing it, I compared closely the
printed version with the original manuscript to settle some discrepancies and ensure accuracy. There are still 
a few minor issues of purely archival interest. 
The printed version uses capital letters $X_i$ for particle coordinates. The symbol
in the hand-written notes is ambiguous, but my personal discussions with Feynnman make me believe that
he intended to use lower-case letters. This is also consistent with later parts of the proof where lower case
letters appear in the printed version. I therefore reverted to lower case $x_i$ in the entire proof. 
I also considered
omitting my own corrections for the sake of originality, but I believe this would have been inappropriate. The version presented here had received Feynman's final approval.

\section{The Proof}

\noindent
Dear Bill,

You asked me to send you what I knew of the conservation laws for a system of $N$ particles on a line all virtually
interacting by a potential $V(ij) \equiv V( x_i - x_j )$ where $x_i$ is the position of particle $i$. If $p_i$ is the
momentum of the $i^{th}$ particle then the total momentum
\be
I_1 = \sum_i p_i
\ee
is conserved. So is the total energy, and that means $\half I_1^2 ~-$ Energy $= I_2$ is conserved
\be
I_2 = \ul{p_i p_j} - \ul{V(ij)}
\ee
where by underlining a term I mean the sum of all distinctly different terms of the form underlined, each taken 
only once, where the indices take all values from $1$ to $N$, except that all indices must be different. [Thus
if $N=3$, $\ul{p_i V(jk)}$ would mean (for $N=3$), $p_1 V(23) + p_2 V(13) + p_3 V(12)$ since $V(ij) = V(ji)$
they are not distinct; while $\ul{p_i p_j p_k}$ is simply the one term $p_1 p_2 p_3$.]

Next, trying to find an $I_3$ that is a constant of the motion (commutes with $I_2$) by trying to start with the
term $\ul{p_i p_j p_k}$ we find, as you discovered,
\be
I_3 = \ul{p_i p_j p_k} - \ul{p_i V(jk)}
\ee
provided $V$ satisfies a special condition (which we call $C_2$):
\be
C_2 ~{\rm{:}}~ \ul{V' (ij) V (jk)} = 0
\ee
which you showed implies $V$ is ${\cal P} (x)$ the Weierstrass Elliptic function ($V' (x) \equiv dV(x) / dx$).

A simple special case is $V = 1/x^2$ and we can consider the general solution as a kind of generalization into complex $x_i$, doubly periodic. (The condition $C_1 ~{\rm :}~ \sum_{i,j} V' (ij) = 0$ is obvious for $V' (12) = - V' (21)$.)

Then it is easy to verify that
\be
I_4 = \ul{p_i p_j p_k p_l} = \ul{p_i p_j V(kl)} + \ul{V(ij) V(kl)}
\ee
is also conserved, as well as an entire string of $I_K$ up to $I_N$.

In this letter I give my proof that $I_2$ commutes with them all and that they all commute with each other. But this
is only true if the potential satisfies beside $C_2$, also a sequence of other conditions
\bea
&& C_3 ~{\rm :}~ \ul { V' (ij) V(jk) V(kl) } = 0\\
&& C_n ~{\rm :}~ \ul{V' (ij) V(jk) \dots V(rs) V(st)} = 0
\eea
(with a $V'$ and $n-1$ $V$'s).

(That $C_3$ is necessary you can see by trying to commute $I_3 , I_4$ directly by hand.) You can imagine my
surprise [that is the demonstration of our lack of deep understanding of these things, - always a surprise!] 
when I checked $C_3$ (for $N=4$ is enough) by writing out the 24 terms for $V= 1/x^2$ which, I think,
implies them also for ${\cal P} (x)$. Obviously each $C_n$ needs only be checked as an identity for $N = n$
particles.

All these $I_n$ are easy to write if we write
\be
I_n = \langle \ul{(p_i+\phi_i)(p_j+\phi_j) \dots (p_m + \phi_m)}\rangle ~~{\rm (}n\rm{\, factors)}
\ee
where $\phi_i \equiv \phi ( x_i )$ and the mean signified by $\langle ~~ \rangle$ is taken on $\phi (x)$ in a gaussian manner so the mean of $\phi$ or any odd number of $\phi$ is zero, and
\be
\langle \phi_i \phi_j \rangle = - V(ij)
\ee
(Omit terms when $i=j$ or formally take $V(ii)=0$.) The mean of $\langle \phi_i \phi_j \phi_k \phi_l \rangle = 
V(ij) V(kl) + V(jk) V(il) + V(ik) V(jl)$. The rule in general is to select any way of combining $\phi$ factor pairs
$\phi_k \phi_l$ and writing a factor $- V(kl$) for each  pair, and adding over all ways of making the selection.

We can deal with all the $I_n$ at once by studying the generating function
\be
t (\mu ) = \left< \prod_{i=1}^N (p_i + \mu + \phi_i ) \right>
\ee
The various $I_n$ are the coefficients of $\mu^{N-n}$ in the power series expansion of $t(\mu )$. Our problem is
then to prove that $t(\mu )$ and $t (\nu )$ commute for any $\mu$, $\nu$. The product $t (\mu ) t(\nu )$ may
be written
\be
\left< \hskip -0.13cm \left< \prod_{i=1}^N (p_i + \mu + \phi_i ) ( p_i + \nu + \chi_i ) \right> \hskip -0.13cm \right>
\ee
where the double mean $\left< \hskip -0.05cm \left< ~~ \right> \hskip -0.05cm \right>$ signifies we take a mean on $\phi$ as described before, 
and on another function $\chi$ with exactly the same rures $\langle \chi_i \chi_j \rangle = - V(ij)$ etc. We consider
this as a double mean on a product of factors $f_i$ which commute with each other for different $i$ and
where (suppressing the index, note $p \phi - \phi p = -i \phi'$)
\be
f = (p + \mu + \phi ) (p + \nu + \chi ) = p^2 + {1 \over 2}(p\phi + \phi p) + {1 \over 2} (p\chi + \chi p) + \nonumber
\ee
\vskip -0.7cm
\be
\mu \chi + \nu \phi + \mu \nu + \phi \chi + {i \over 2} ( \phi' - \chi' )
\ee
For the $t(\nu ) t(\mu )$ we reverse $\mu$ and $\nu$. For convenience we also reverse $\phi$, $\chi$ and the
reversed factor is the same except for the last term which becomes $-{i \over 2} (\phi' - \chi' )$.
The commutator is the difference and this is a term proportional to $i$ (written {\it Im})
\vskip -0.8cm
\be
[t(\mu ) , t(\nu ) ] = 2 {\it Im} \left<\hskip -0.13cm \left< \prod_i p_i^2 + \mu \nu + {1 \over 2} [p_i \phi_i + \phi_i p_i ] + 
{1\over 2} [p_i \chi_i + \chi_i p_i ] \right. \right. \nonumber
\ee
\vskip -0.5cm
\be
~~~~~~~~~~~~~~~~~~~~\left. \left. + \, \mu \chi_i + \nu \phi_i + \phi \chi_i + {i \over 2} (\phi'_i - \chi'_i ) \right>\hskip -0.1cm \right>
\ee
\vfill
\eject
\noindent
There must be an odd number of $i (\phi' - \chi' )$ factors, at least one, so let us start with one of these, 
say for the index $i$. Begin with the mean of $i \phi'_i$ times all the other factors $f_j$ for $j \neq i$.
In taking the mean on $\phi$ we get zero unless we connect $\phi'_i$ with a $\phi$ from some other factor,
say $f_j$. In each factor $\phi$ appears linearly, either alone or ``linked", multiplied by a $\chi$ in $\phi_j \chi_j$.
Let us take the case it is linked, when we get $V' (ij) \chi_j$ on taking the mean. Now the $\chi_j$ on taking the mean
must also vanish or find a $\chi$ in another factor $f_k$ to connect to. Suppose that is also linked. Then we are
up to $i V' (ij) V(jk) \phi_k$. Again we must find $\phi_l$ in $f_l$ to connect to make a $V(kl)$ factor, etc., 
for say $n$ linked terms. Finally in some term, say $m$, we choose a linear unlinked factor and have a term with
a factor
\be
i V' (ij) V(jk) \dots V(.\, m) \cdot (-1)^n
\ee
times either (a) a $p_m$ (acting symmetrically before and after $x_m$) or something not dependent on $m$, i.e., 
(b) $\nu$ if we end on a $\phi$ ($n$ odd) or $\mu$ if on a $\chi$ ($n$ even). \hskip -0.08cm (c) with the last factor being
$V' (. \, m)$ and $+{i \over 2}$ for $n$ even, $-{i\over 2}$ for $n$ odd.

Now had we started instead with the $-i\chi'_i$ term we can follow exactly the corresponding sequence as above
for $i\phi'_i$ for linked and non linked factors {obtaining the same final factor \hskip -0.06cm (14) \hskip -0.05cm with opposite sign, \hskip -0.03cm and for the
three cases:}

(a) the same $p_m$ so the result cancels

(b) here $\mu$ comes if $n$ is odd, and $\nu$ if $n$ even so in sum we get $(-1)^n (\mu - \nu )$ times (14).

(c) the factor is real, and we have used up two $(\phi' - \chi' )$ factors.

For given $n$ in (14) we have used up all the factors $f_i, f_j \dots f_m$ and the remaining product of $f$'s involves
all indices other than these. Now consider all the path structures that are linked with the same $n$ factors but in a
different order. Summing over all these in the case (b) we get a factor ($\mu - \nu$ times)
\be
C_n = \ul{V' (ij) V(jk) \dots V(.\, m)} ~~[n ~\rm{factors}]
\ee
which we shall later show vanishes for our special potentials. Finally in case (c) we have used up an even number of
$i(\phi' - \chi' )$ factors, but in the commutator there must be an odd number so we have at least one left. We can
start with it and proceed to argue just, as before, with the expression with $N-n$ factors, arguing recursively.
Only case (c) survives again, but again there is an odd number left, until there is only one left, in which case it cannot
terminate in case (c), but only (a) cancelling  directly or (b) which vanishes because  of (15).

Therefore all the $I_n$ are constants of the motion and commute with one another.

To finish the demonstration we shall have to prove that $V$ satisfies all the conditions. We assume $C_2$ of course,
and then by induction, proving $C_n$ supposing $C_k$ true for $k=2$ to $k-1$. Concentrate on the case
$V(12) = 1/(x_1 - x_2 )^2$ first. Consider the expression (15) as a function of $x_1$. There are apparently
many singularities, double poles from $V(ij)$ and triple from $V' (ij)$. We show, in fact, there are no poles
and, since the function is independent of $x_1$ as $x_1 \to \infty$ it is independent of $x_1$. (We can do this for
any variable so it depends on none, and being 0 as all approach $\infty$ it must be zero.)

The singularities come when $x_1$ gets near another $x_j$. We choose a particular other particle 2 (i.e., $j=2$)
and show there is no singularity when $x_1$ approaches $x_2$. Obviously we need only study terms involving
$V' (12)$ and $V (12)$ for no other term is singular when $x_1 \approx x_2$. The term proportional to $V' (12)$
from (15) is ($i=1, j=2$ and $i=2,j=1$; $V' (12) = - V' (21)$)
\be
V' (12) \, \ul{ [V(2k) - V(1k)] \, V(kl) \dots V(.\, m)}
\ee
We see, although $V' (12)$ is a triple pole, the coefficient reduces it for it vanishes as $x_2 \to x_1$. Using $C_2$
the first two factors (before $V(kl) \dots V(.\, m)$) can  be written as
\be
-V' (2k) [ V(k1) - V(21) ] - V' (k1) [ V(12) - V(k2) ]
\ee
The terms not containing $V(12)$ produce no pole, our pole comes from the coefficient of $V(12)$ which is
\be
{\rm{Coeff~of~}} V(12) = \ul{V' (1k) V(kl) \dots V(. \, m)}
\ee
plus a similatr term with $1$ replaced by $2$ (i.e., $V' (2k)$ in place of $V' (1k)$. Now there are $n-1$ factors in this
expression so we can use our assumed $C_{n-1}$ to rearrange it. In the sum $C_{n-1}$ the variable $2$ appears
not only in the first position but in every other position as well, where $k,l,\dots m$ appear. The sum is zero so (18)
is the negative of the sum for all these replacements. We write only the replacement at $l$ as an example, and
a $\sigma$ to mean sum all other positions (except the very first).
\bea
\rm{Coeff~of~} V(12) = &{\hskip-0.2cm -\, \sigma V'(kl) V(l\, .) \dots V(.\, 2) V(2 \, .) \dots V(. \, m)} \nonumber \\
 &{\hskip -0.2cm -\, \sigma V'(kl) V(l\, .) \dots V(.\, 1) V(1 \, .) \dots V(. \, m)}
\eea
(terms contain the variable twice except the last one where it appears once).
Next we must add all the contributions from $C_n$ in (18) which contain a factor $V(12)$. These come because
some pair like $jk$ or $kl$ are $12$, or are $21$. The Coeff of $V(12)$ coming from this is
\bea
&+ \, \sigma V' (kl) V(l\, .) \dots V(.\, 1) V(2\, .) \dots V(.\, m) \nonumber \\
&+ \, \sigma V' (kl) V(l\, .) \dots V(.\, 2) V(2\, .) \dots V(.\, m)
\eea
(all terms have both $1$ and $2$ except if the $V(12)$ came from the last factor in (16), in which case $m=1$
or $2$ alone). Thus combining (21) and (20) (the terms where $m=1$ or $2$ cancel) and reinstating the $V(12)$
the putative pole in $C_n$ is
\bea
V(12)\hskip -0.6cm && [ V' (kl) V(l\, .) \dots V(.\, 2) - V' (kl) V(l\, .) \dots V(.\, 1) ] \cdot  \nonumber \\
&& [V(1\, .) \dots V(.\, m) - V(2\, .) \dots V(.\, m)]
\eea
But each factor vanishes as $x_2 \to x_1$ as $(x_2 - x_1 )$ so two such factors are supplied to cancel the
expected singularity, QED.
I think the proof can be extended to when $V$ is Weierstrass' Elliptic function $\cal P$ for the algebra is
immediately extended to complex variables $z_i$ for $x_i$. Then if $V(12)$ has a pole because $z_2 - z_1 =$
a period of $\cal P$ each of the factors also vanish for they are all singularly periodic functions.
(A constant $\gamma$ may be added to $V$ in $C_n$ for the coefficient of $\gamma^k$ is $C_{n-k}$
which we assume to vanish.)

Well, that is what I promised you. I am disappointed in the proof for two reasons. First it requires two parts
and probably there is a way to do them both at once, but second I learn nothing, no real clue as to why all
this works, and what it means.

Thank you very much for your papers and reprints, which have arrived, but which I haven't studied yet.

\section{Epilogue}

After Feynman completed this proof we continued working on the model. Part of this work consisted in introducing
extended dynamical quantities, generalizing $I_n$, that involved both coordinates and momenta in their leading term.
These closed
under commutation into what is nowadays called the $W_\infty$ algebra (although Feynman didn't care for ``algebras") and their time evolution was linear in lower-order quantities, eventually closing with $I_n$, which afforded a detailed probing of the model's dynamics. This and some
other related work was also reported by Feynman in subsequent letters to Sutherland but remains otherwise unpublished.
(The manuscripts are in my archive.)

My involvement ended in August 1987 when I left Caltech for my first postoctoral position. Later that fall Feynman
suffered another relapse and underwent aggressive treatment, which made him weak and reduced the time he
could spend in the office. I visited Caltech during the winter break of 1987-88 and talked with Feynman one last time.
He expressed intense interest in the interpretation of the Calogero model as particles with fractional statistics but his failing health
kept him from making any additional contributions.

He passed on February 15, 1988.

\vskip 0.3cm
{\it {\ul {Acknowledgements}}:} I am thankful to Hiroshi Ooguri for contacting me about archival material of Feynman, which prompted me to revisit my files. Elements of the story narrated here have been presented in colloquia delivered at ICTS Bangalore, the City College of New York, Stony Brook and LPTMS Orsay. The encouragement of several people attending these lectures has been the main catalyst in motivating me
to prepare this account.


\large{\bf Appendix: Feynman's manuscript -- the first page}
\vfill
\eject
\pagestyle{empty}
\begin{figure}
\vskip -3.5cm \hskip -3.6cm \includegraphics[scale=0.97]{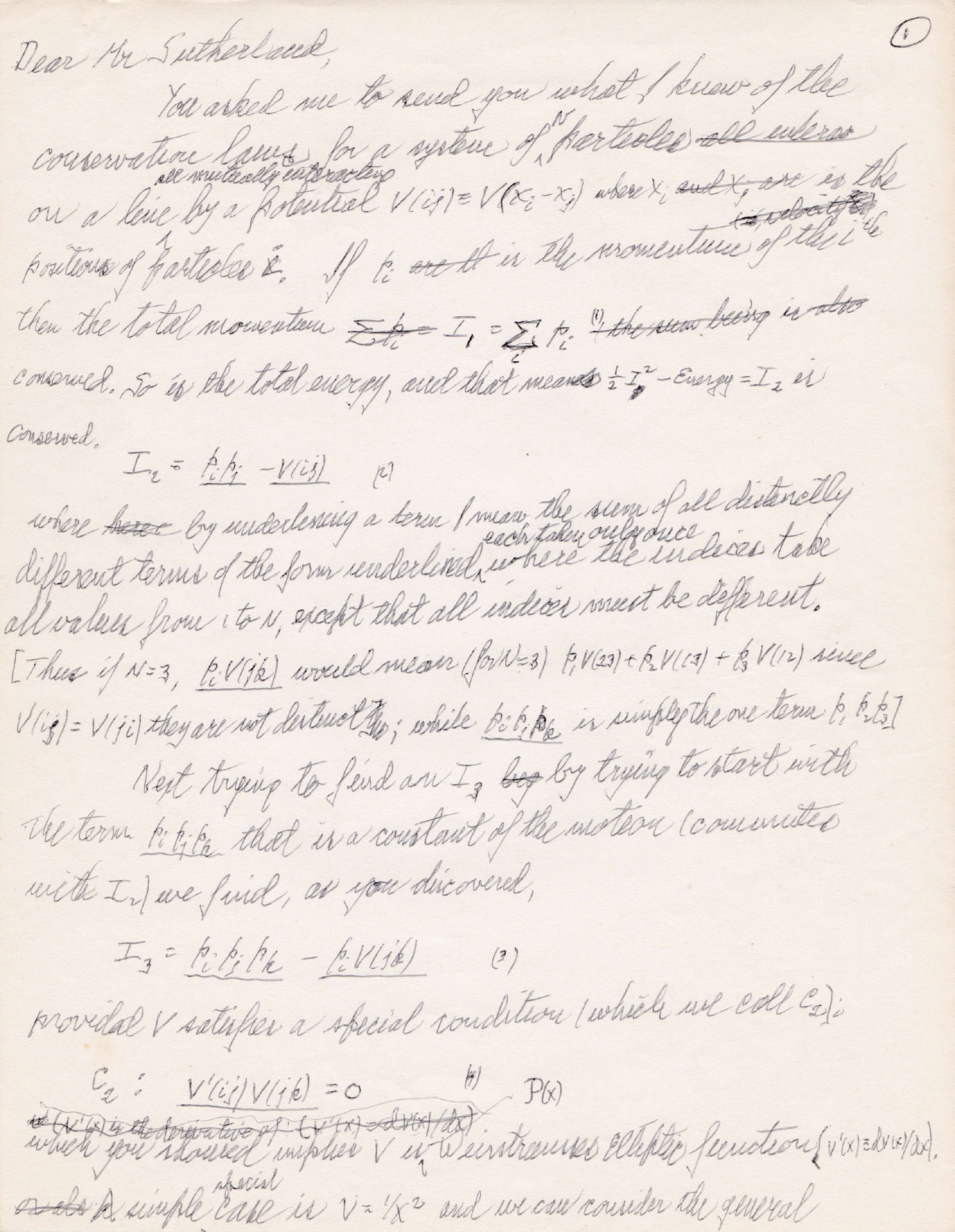} 
\end{figure}

\end{document}